**Pressure variation of Luttinger liquids parameters in single wall carbon nanotubes networks**

*Short title* : **Luttinger liquid networks under pressure**

F. Morales (*), M. Monteverde (†), and M. Núñez-Regueiro

Centre de Recherches sur les Très Basses Températures, C.N.R.S., BP 166, 38042 Grenoble, France

We measure electrical transport on networks of single wall nanotube of different origin as a function of temperature $T$, voltage $V$ and pressure $P$. We observe Luttinger liquid (LL) behavior, a conductance $\propto T^{\alpha}$ and a dynamic conductance $\propto V^{\alpha}$. We observe a sample dependent $P$ variation of the $\alpha$ parameters, interpreted as fermi level changes due to pressure induced charge transfer. We show how, through standard four-leads and crossed configuration methods, it is possible to determine $\alpha_{bulk}$ and $\alpha_{end}$, respectively. We study and discuss the pressure and doping level dependences of the number of channels $N$, the LL parameter $g$ and the intra-rope tube-tube coupling constant $U$.

**PACS** 61.46.Fg, 73.63.Fg, 62.50.+p

---

(*) On sabbatical leave from Instituto de Investigaciones en Materiales, Universidad Nacional Autónoma de México. A. Postal. 70-360. México, D. F. 04510 Mexico.

(†) Present address Service de Physique de l'Etat Condensé (CNRS URA 2464), DSM/DRECAM/SPEC, CEASaclay, 91191 Gif sur Yvette Cedex, France

It was recently found[1]that a network of single wall carbon nanotubes (SWNT) behaves as a Luttinger liquid (LL) in transport measurements under pressure. In the same way as was reported for individually contacted SWNT rope samples, power laws in temperature $T$ and voltage $V$ were measured, i.e. a conductance $\propto T^{\alpha}$ and a dynamic conductance $\propto V^{\alpha}$. The exponent obtained corresponds to $\alpha = 2\alpha_{bulk} = \alpha_{bulk-bulk}$, i.e. the data correspond to a bulk-bulk contacted array of SWNT. However, for a complete interpretation of the LL properties, a determination of $\alpha_{end}$, the end exponent, is necessary. We show here a more detailed interpretation of the relation between a single tube and our macroscopic network samples $\alpha$ exponents and the way in which the end exponent can be obtained from a network of SWNT. Electrical resistivity measurements rendering both the conductance $G$ as a function of temperature $T$ and the dynamical conductance $dI/dV$ as a function of voltage $V$ and temperature $T$ were performed at pressures up to 12 to 23GPa, on several different samples. We use the obtained values to analyze details of the origin of the pressure variation of the exponents.

Samples of type A were produced with a method that ensured a reducing environment favoring electron doping of the nanotubes. The SWNT were obtained from MER Corporation, they have average diameter of 0.7-3 nm, and lengths from 2 – 20 $\mu m$. In order to prepare the samples for transport high pressure studies they were purified by heating in air to 300 °C for 24 h and dispersed by ultrasound in hydrochloric acid, and refluxed for 4 h at 100 °C. After that, they were washed with distilled water and ethanol (95 % spectrophotometric grade) and filtrated through a membrane filter of 0.20 $\mu m$ pore size. This method [2] produces a mat-like sample that was dried at 100 °C for about 2 h. The bucky-paper was compacted previous to insertion in the pressure cell. Samples of type B are from the same origin as those used in Ref. [1], i.e. untreated and unpurified, though air exposed, SWNT bundles. The electrical resistance measurements were performed in a sintered diamond Bridgman anvil apparatus using a pyrophillite gasket and two steatite disks as the pressure medium[3]. The Cu-Be device that locked the anvils does not allow measurements releasing the pressure and could be cycled between 4.2K and 300K in a sealed dewar.

In ref. [1] it was found that conventional four lead (4W) electrical resistivity measurements furnished evidence for a bulk-bulk contacted network of nanotubes, showing the behavior expected from such a Luttinger liquid network with an average exponent $\alpha$. We show now how this average is related to the individual exponent $\alpha$ of one single bulk-bulk nanotube contact. In conductivity measurements on individual MWNT [4] and on individual ropes of SWNT [5, 6], low quality contacts are usually unavoidable, so what is measured with those devices is the tunnel conductivity of two NT-leads junctions in series. While measurements of 4 wires (or 2 wire) conductivity on a macroscopic network of NT (MWNT or SWNT ropes) involve a statistic average of the tunnel conductivity of both NT-leads and, mainly, of a large number of NT-NT junctions. The conductivity of each element of the junctions network (JN) is expressed in equation (1)

$$G_i = G_{0i} E^{\alpha_i} \quad (1)$$

where $G_i$ is the conductance of the $i$ junction, for a LL we expect a power law of the relevant energy $E$ ($T$ if $eV << k_B T$ or $V$ if $eV >> k_B T$) and exponent $\alpha_i$ which depends on the type of junction being tested [7]. The conductivity of the JN is obtained by adding junctions both in parallel and in series. If we assume that all junctions are of the same type and consequently have the same exponent $\alpha_i = \alpha_0$, then the JN conductance is straightforward and behaves also as a power law with an exponent identical to the one corresponding at the involved junction ( $\alpha_0$ ).

Now, we can consider some inhomogeneity between the involved elements and therefore between theirs exponents. We consider the inhomogeneity in the $\alpha$ exponent as a square distribution of width $2\Delta$ around $\alpha_0$. In this case, the addition of the JN can be expressed in expression (2)(for parallel addition)

$$G_{parallel\,(\alpha_0,\Delta,E)} \propto \int_{\alpha_0-\Delta}^{\alpha_0+\Delta} E^{\alpha} d\alpha = \frac{E^{\alpha_0}}{\log(E)}\left(E^{\Delta} - E^{-\Delta}\right) \qquad (2)$$

It is evident that this expression doesn't behave as a clean power law. But, we can easily estimate how far is by taking the logarithmic derivative (3) of expression (2). In this way we can study the local $\alpha$ exponent ( $E$ dependent), defined as $\alpha_{l(\alpha_0,\Delta,E)}$.

$$\alpha_{l(\alpha_0,\Delta,E)} = \frac{\log(G_{(\alpha_0,\Delta,E+dE)}) - \log(G_{(\alpha_0,\Delta,E)})}{\log(E+dE) - \log(E)} \qquad [3]$$

Considering typical experimental values of both temperature and voltage, as well as exponent $\alpha_0$ and inhomogeneity (taken here as $2\Delta = 30\%\alpha_0$), we can estimate $\alpha_{l(\alpha_0,\Delta,E)}$. The largest variation of this local exponent, within experimental $\alpha_0$, $\Delta$ and $E$ parameters ranges, is a measure of how far is expression (2) from a clean power law (note that expression (3) for negatives $\alpha_0$ actually corresponds to series addition case). This variation is less than 0.5%, which means that for all practical purpose the observed behavior is a clean power law.

Another consideration that must be evaluated is how far is the average $\alpha_{I(\alpha_0,\Delta,E)}$ (the exponent that should be observed in the JN experiment) from the center of the inhomogeneous distribution, $\alpha_0$. Using again expression (3), we obtain a deviation between the measured exponent and $\alpha_0$ of less than 1%.

So, 4 wires and 2 wires experiments over a network of LL tunnel junctions represent a statistical accurate measure of NT-NT tunnel conductivity.

However, for a deeper analysis of the properties of SWNT it is convenient to measure also the power law coefficient corresponding to an end contact, in order to estimate the number of conducting channels $N$, the value of the $U$ parameter and of the crucial $g$ parameter. A crossed configuration [8] as shown in the inset of Fig 2, is a suitable geometry to obtain a lead-nanotube end contact and to be able to measure the power laws corresponding to this configuration. The lead across the sample crushes down the nanotubes below it, and generates an "end" contact in the middle of the sample. Furthermore, by measuring in the crossed configuration we measure exclusively the lead-nanotube junction.

We show on Fig. 1 the measurements obtained from the four-lead configuration and in Fig. 2 those obtained for the crossed measurements on a typical sample of type A. Samples of type B give for the 4W configuration results identical as those of Ref. [1]. We observe that the crossed configuration yields almost perfect power laws in all the temperature range, while the four lead measurements have a slight negative curvature, that can be attributed to the fact that in this configuration we measure a network of junctions plus probably the internal resistance of the nanotubes. The main differences between both type of samples is that pressure increases $\alpha$ for the A-type, while it decreases $\alpha$ for the B type. This can be understood if we consider that exposure to air induces oxygen adsorption and that pressure increases the oxygen-carbon charge transfer [9,10]. The reducing condition during the preparation of samples A, makes us expect an original electron doping of these samples that is gradually compensated by the oxygen pressure charge transfer.

From the power law $G(T) \sim T^{\alpha}$ of both configurations, we obtain the exponents [7] $\alpha_{4W}/2 = \alpha_{b-b}/2 = \alpha_b = (1/g + g - 2)/8N$ and $\alpha_X = \alpha_e = (1/g - 1)/4N$. These exponents coincide with similar power law behaviors of $dI/dV \sim V^{\alpha}$ that we have measured at constant temperatures, similar to what be reported done in Ref . [1]. From these two exponents the number of conducting channels is uniquely determined eliminating $g$ using expression (4).

$$N = \frac{\alpha_b}{2\alpha_e^{\,2} - 4\alpha_e\alpha_b} \qquad (4)$$

We obtain a neat decrease (increase) with pressure of $N$ for samples of type A (B). As described above, an increase of charge transfer(from adsorbed oxygen) due to increasing pressure can explain our results if samples of type A (B) are originally electron (hole) doped. Samples of type A would thus increase their Fermi level with pressure, emptying previously electron occupied channels, while samples of type B would increase their hole occupation and reduce their Fermi level with pressure. We can try to quantify this assertion.

In SWNTs the number of conducting channels ( $N$ ) not only strongly depends on Fermi level, but also on its diameter and chirality. As in this work we measured macroscopic amounts of SWNT, we will consider an average of the densities of state corresponding to all possible chiralities of metallic SWNT between experimental samples diameters, according to a square distribution (for simplicity). In this way we obtain a theoretically average $N$ as function of band filling $n$. On the other hand, depending on the preparation the SWNT have, at ambient pressure, either an electron or hole doping. Besides, air exposure while mounting induces additional oxygen adsorption of both type of samples. So we calculate the average doping of all considered SWNT for each doping level $n$, in order to obtain the number of conducting channels as a function of doping. This $N(n)$ dependence can be easily compared with the measurements performed on this work over different samples. It must be assumed for each sample a different starting doping level at ambient pressure ( $n_0$) and a different pressure charge transfer between oxygen and SWNT rate ( $dn/dP$), determined by the type and amount of the dopants of each sample. The adjustment between calculation and measurements can be seen on Fig. 3 for two typical samples of type A (U67F) and B (S80). Reasonably, we must consider different charge transfer rates depending on the initial state, A or B. We observe a very good agreement between the number of channels $N$ obtained from our measurements (by means of expression (4) and the $N$ calculated (statistical average over different diameters and chiralities of the density of states) within an oxygen charge transfer under pressure hypothesis.

One way to verify if we are in fact measuring $\alpha_e$ by means of the crossed configuration is to calculate the ratio the ration $\alpha_e/\alpha_b$, whose variation with doping level we show on Fig. 4. We observe that the latter ratio is almost constant with pressure within the experimental error. It is around 2.4, that is the expected ratio for a LL system, as all other theories yielding similar power laws yield a strictly $=2$ ratio [11,12], confirming that the crossed configuration furnishes the end exponent. We can also determine uniquely the $g$ parameter, shown on Fig. 4(middle panel). Its variation with doping is more important, with a tendency to increase with hole doping.

Egger[13] has calculated the $g$ parameter expected from $N$ interacting channels as function of a dimensionless coupling constant $U$ that depends on the size and dielectric constant of the nanotubes. He obtains $g=(1+N\cdot U)^{-1/2}$, from where we can extract the variation of $U$ with doping level (Fig. 4 upper panel). Within the large errors accumulated by our calculations, we obtain values similar to those estimated by Egger, with a tendency for a decrease of $U$ with the number of occupied channels, symmetric with respect to doping, i.e. as the $g$ parameter is almost constant the increase of $N$ is compensated by the decrease of $U$.

In conclusion, we have described how macroscopic samples of SWNT behave as a network of bulk-bulk connected SWNT junctions with an average $\alpha$ parameter. We have shown how the parameter corresponding to an end contact, $\alpha_e$, can be measured using a crossed configuration in such macroscopic samples. By measuring both the bulk and the end exponent we are able to uniquely determine the number of occupied channels, $N$ the $g$ parameter and their ratio. By analyzing the variation of $N$ against an average density of states, we find strong support for a charge transfer explanation of the variation of $N$ with pressure, i.e. pressure increases the hole transfer from adsorbed oxygen molecules.

We thank J.A. Azamar and R. Escudero for providing the type A samples and S. Tahir and P. Bernier for furnishing us with the type B samples. We gratefully acknowledge discussions with R. Egger. F.M. acknowledges support by DGAPA-UNAM.

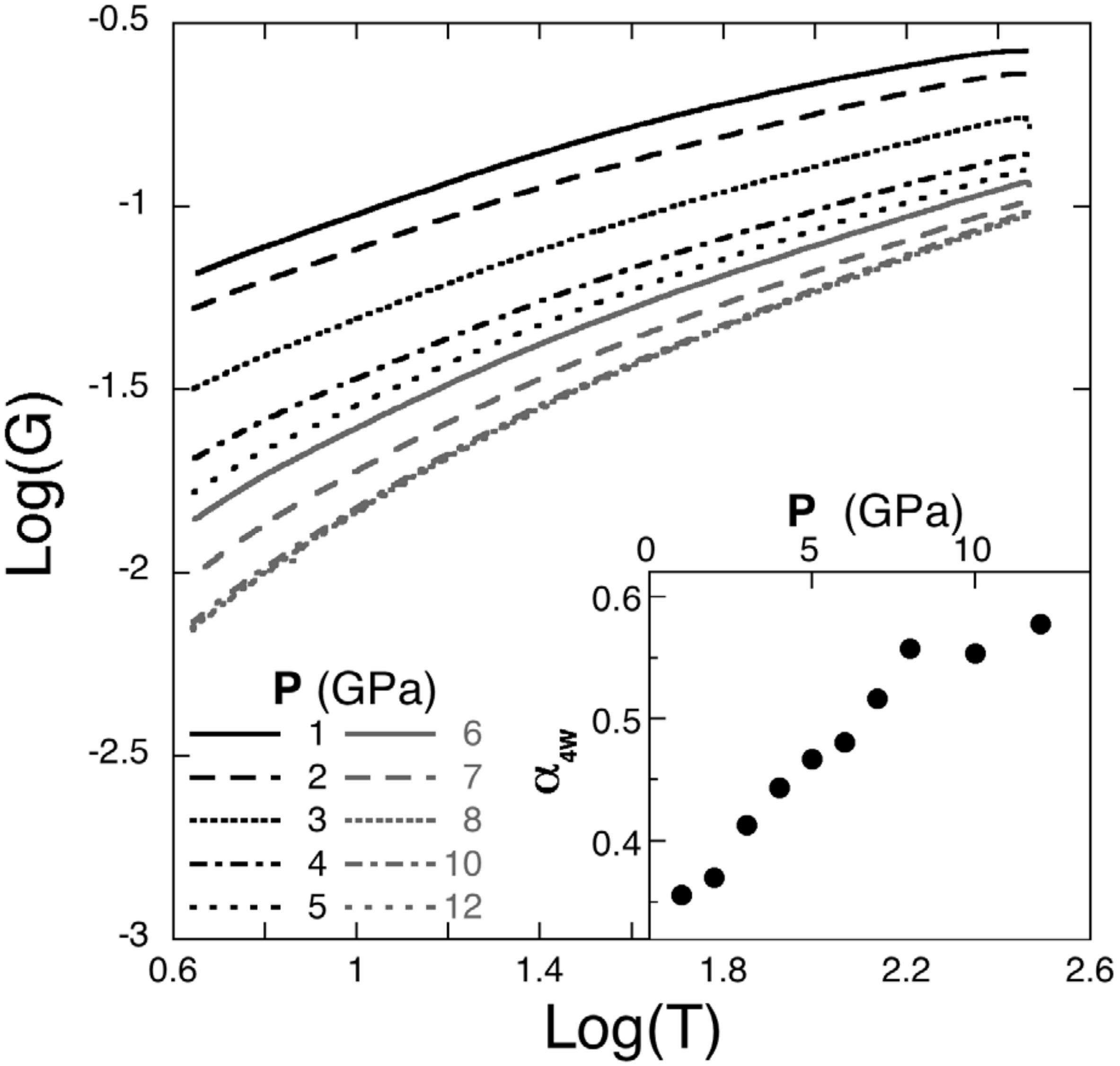

**Figure 1 Typical temperature dependence of the four leads conductance of a sample U67F (type A, purified samples) for different pressures. We observe a power law behavior at low temperatures. Inset : pressure dependence of the** $\alpha_{4W} = \alpha_{bulk-bulk}$ **exponent.**

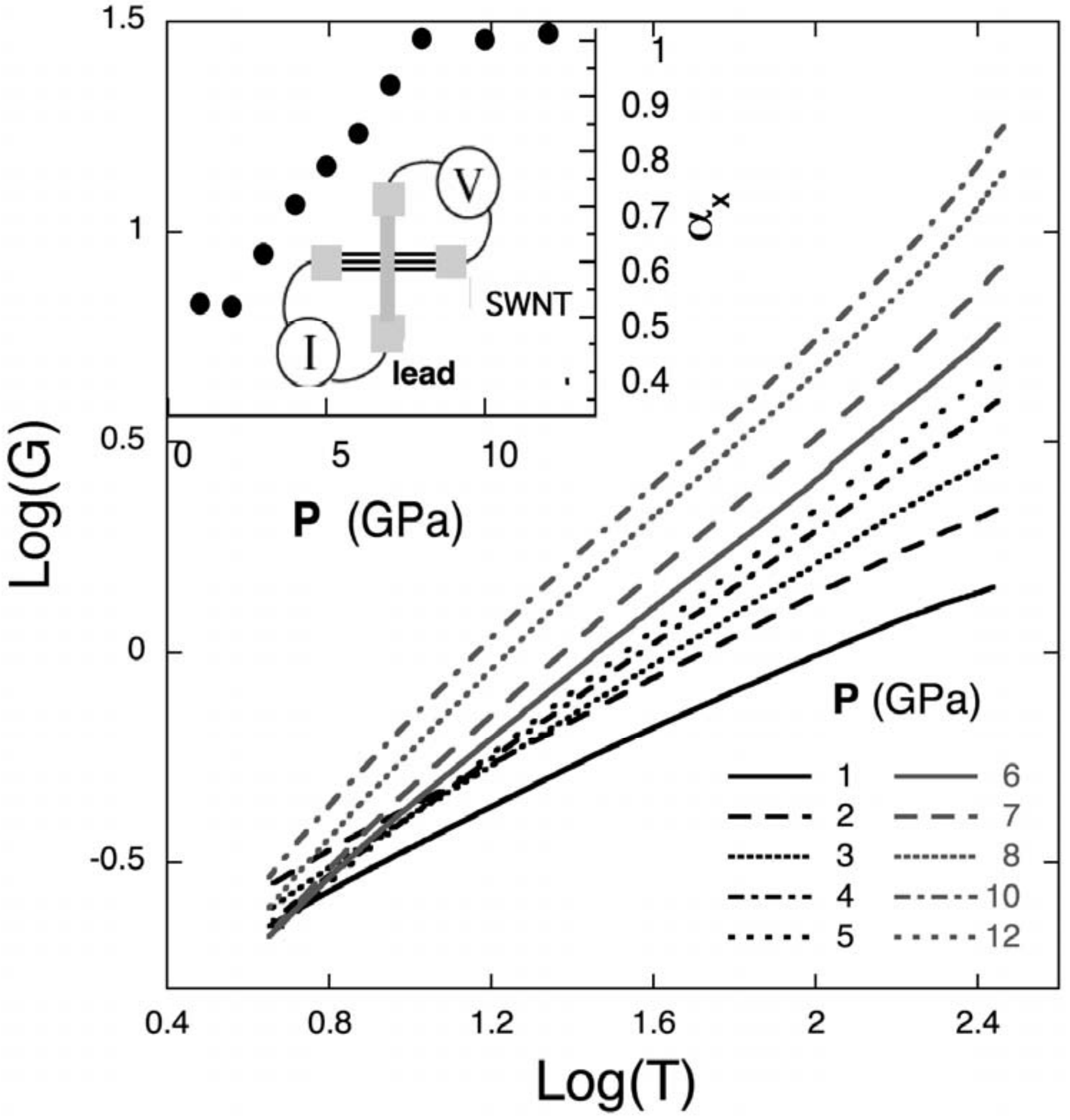

**Figure 2 Typical temperature dependence of the crossed leads conductance of a sample U67F (type A, purified samples) for different pressures. We observe a power law behavior at all the temperature range with an exponent $\alpha_X = \alpha_{end}$. Inset: variation of $\alpha_X$ with pressure and diagram showing the crossed configuration.**

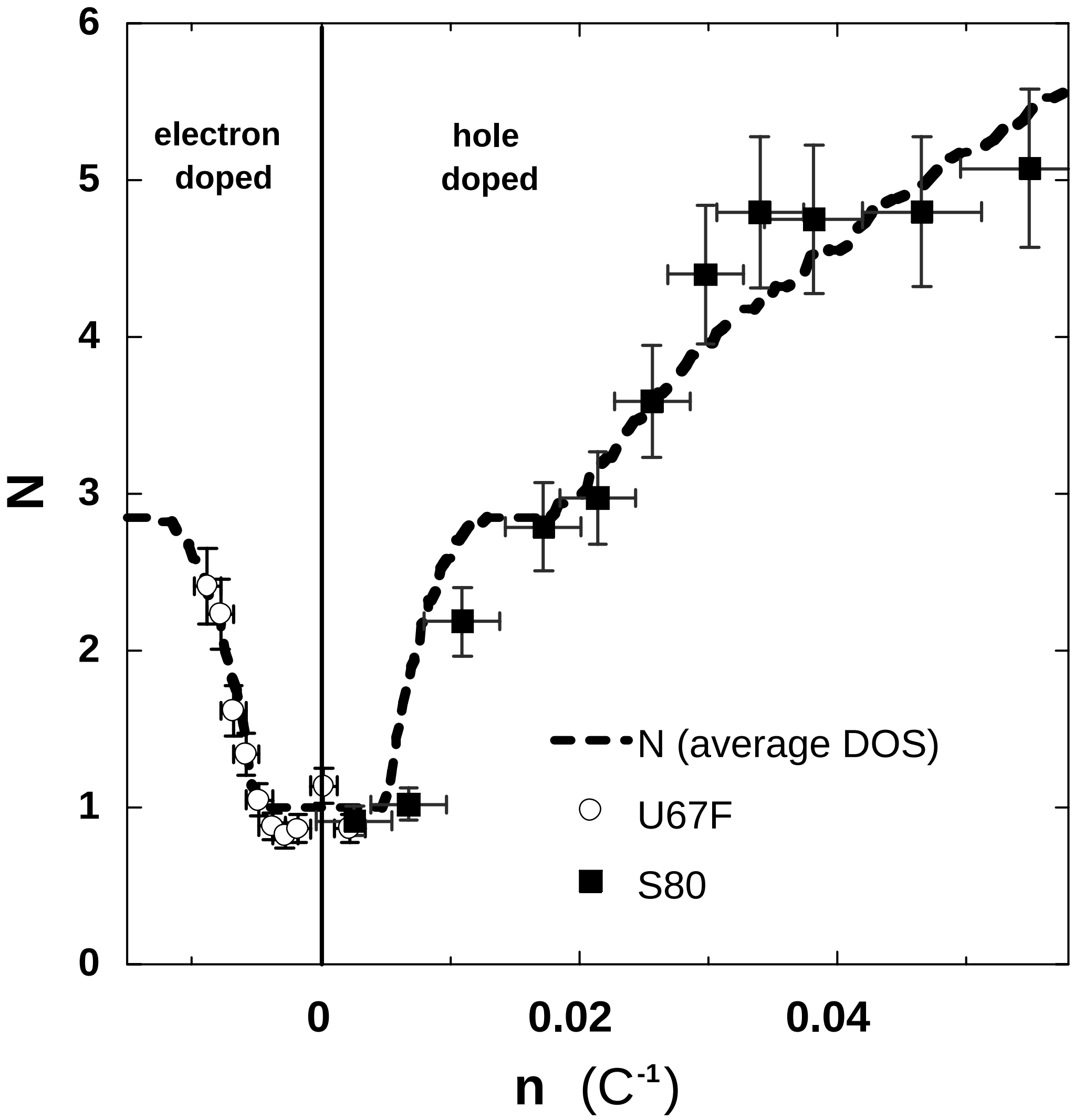

**Figure 3 Number of conducting channels as function of pressure-induced hole-doping. Sample S80 (of type B, as grown, not purified samples) has a doping rate of** $dn/dP|_{S80} = 0.0042 \pm 0.0003\,\text{holes}\ C^{-1}GPa^{-1}$ **with an initial doping of** $n_0|_{S80} = 0.001 \pm 0.001\ C^{-1}$ **(hole doped). Sample U67F (of type A, purified samples) has a doping rate of** $dn/dP|_{U67F} = 0.0010 \pm 0.0001\,\text{holes}\ C^{-1}GPa^{-1}$**, and an initial doping of** $n_0|_{U67F} = -0.010 \pm 0.001\ C^{-1}$ **(electron doped). The dashed thick line corresponds to the**

**statistical average of the number of conducting channels considering all metallic tubes within the diameter distribution of the samples (see text).**

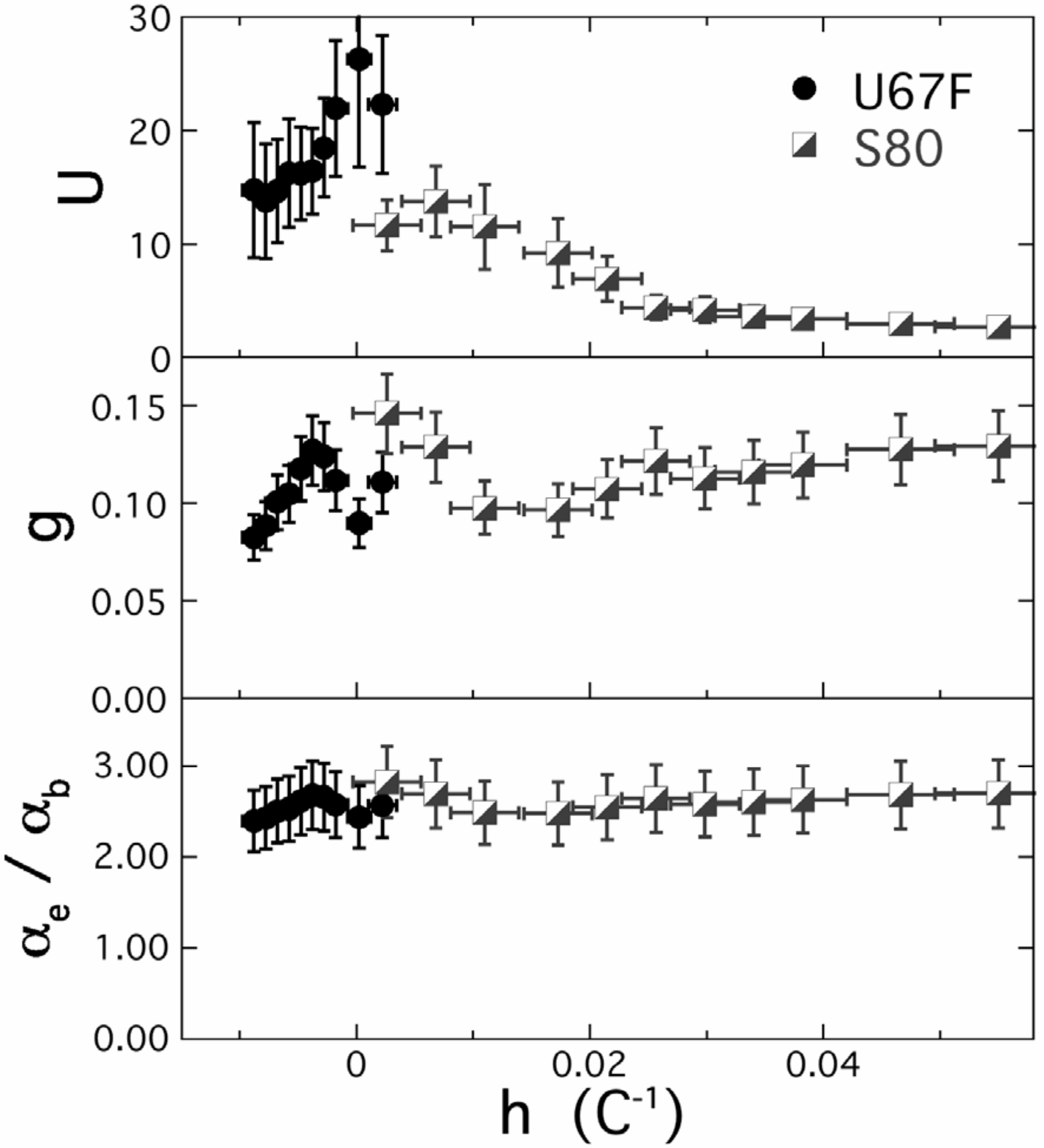

**Figure 4 Variation of the tube-tube intra-rope coupling $U$, the LL parameter $g$ and the ratio $\alpha_e / \alpha_b$ as a function of hole doping**